\RequirePackage[hyphens]{url}
\documentclass[final,5p,times,twocolumn]{elsarticle}
\usepackage{lineno,hyperref}
\usepackage{amsmath}
\usepackage{siunitx}

\usepackage{array}
\usepackage{booktabs}

\usepackage[english]{babel}
\usepackage[autostyle]{csquotes}
\usepackage{float}
\usepackage{blindtext}
\usepackage{mathptmx}
\usepackage{graphicx}
\usepackage{subfig}
\usepackage{tabularx}
\usepackage{fleqn}
\usepackage{natbib}
\usepackage{newcent}  % font, Century Schoolbook
\usepackage{amssymb}  % math
\usepackage{amsthm}   % theorems
\usepackage{epigraph} % quotations
\usepackage{booktabs} % nice looking tables
\usepackage{fancyhdr} % fancy headers and footers
\usepackage{color}    % allow colors
\usepackage{pdfpages} % include assignment PDFs
\usepackage{texshade} % sequence alignments
\usepackage[usenames,dvipsnames]{xcolor}
\usepackage[titletoc,title]{appendix}
\usepackage[dotinlabels]{titletoc}
\usepackage{lscape}
\usepackage{pdflscape}
\usepackage{gensymb}
\usepackage{rotating} % Provides {sideways}{sidewaysfigure}{sidewaystable} environments
\usepackage{textcomp}
\usepackage{float}
\usepackage{hyperref}
\usepackage[hyphenbreaks]{breakurl}

\usepackage[hyphens]{url}
\usepackage{breakcites}

 \biboptions{super,sort&compress }
\journal{Journal of Alloys and Compounds}

\begin{document}

%\preprint{APS/123-QED}
\begin{frontmatter}
\title{\textit{Metastable Phase Diagram and Precipitation Kinetics of Magnetic Nanocrystals in FINEMET Alloys}}% Force line breaks with \\
%% Group authors per affiliation:

%% Group authors per affiliation:
\author[label1]{Rajesh Jha\corref{mycorrespondingauthor}}
\ead{rajeshjha@mines.edu}
\author[label2]{David R. Diercks}
\author[label1]{Aaron P. Stebner}
%% or include affiliations in footnotes:
%\author[label1,label4]{Cristian V. Ciobanu\corref{mycorrespondingauthor}}
\author[label1]{Cristian V. Ciobanu\corref{mycorrespondingauthor}}
%\ead[url]{http://inside.mines.edu/~cciobanu/}
\ead{cciobanu@mines.edu}
%\author[label4]{Cristian V. Ciobanu\corref{mycorrespondingauthor}}
\cortext[mycorrespondingauthor]{Corresponding author}
%\ead{cciobanu@mines.edu}
%\ead{rajeshjha@mines.edu}

\address[label1]{Department of Mechanical Engineering, Colorado School of Mines, Golden, Colorado 80401, USA}
\address[label2]{Department of Metallurgical and Materials Engineering, Colorado School of Mines, Golden, Colorado 80401, USA}
%\address[label4]{Department of Mechanical Engineering and Materials Science Program, 1610 Illinois Street, Brown Hall, W470A, Golden, Colorado, USA}

\begin{abstract}
Research over the years has shown that the formation of the Fe$_3$Si phase in FINEMET (Fe-Si-Nb-B-Cu) alloys leads to superior soft magnetic properties.
In this work, we use a CALPHAD approach to derive Fe-Si phase diagrams to identify the composition-temperature domain where the Fe$_3$Si phase can be stabilized.
Thereafter, we have developed a precipitation model capable of simulating the nucleation and growth of Fe$_3$Si nanocrystals via Langer-Schwartz theory. For optimum magnetic properties, prior work suggests that it is desirable to precipitate Fe$_3$Si nanocrystals with 10-15 nm diameter and with the crystalline volume fraction of about 70 \%. Based on our parameterized model, we simulated the nucleation and growth of Fe$_3$Si nanocrystals by isothermal annealing of Fe$_{72.89}$Si$_{16.21}$B$_{6.90}$Nb$_{3}$Cu$_{1}$ (composition in atomic \%). In numerical experiments, the alloys were annealed at a series of temperatures from 490 to $\SI{550}{\degreeCelsius}$ for two hours to study the effect of holding time on mean radius, volume fraction, size distribution, nucleation rate, number density, and driving force for the growth of Fe$_3$Si nanocrystals. With increasing annealing temperature, the mean radius of Fe$_3$Si nanocrystals increases, while the volume fraction decreases.
We have also studied the effect of composition variations on the nucleation and growth of Fe$_3$Si nanocrystals.
As Fe content decreases, it is possible to achieve the desired mean radius and volume fraction within one hour holding time. The CALPHAD approach presented here can provide efficient exploration of the nanocrystalline morphology for most FINEMET systems, for cases in which the optimization of one or more material properties or process variables are desired.
\end{abstract}

\begin{keyword}
\texttt{Soft magnetic alloys \sep FINEMET \sep CALPHAD \sep Thermocalc \sep TTT Diagram \sep Fe$_3$Si ($\alphaα^{''}$-(Fe, Si) D03) phase }
%\MSC[2010] 00-01\sep  99-00
\end{keyword}

\end{frontmatter}

%\maketitle

%\tableofcontents

\section{INTRODUCTION}\label{introduction}

FINEMET is a soft magnetic material based on the Fe-Si-Nb-Cu-B system,  developed by Yoshizawa and his group at Hitachi in 1988 \cite{yoshizawa1988new,FINEMET_HITACHI}. Due to their high saturation magnetic flux density \cite{yoshizawa1988new, FINEMET_HITACHI}, low core losses \cite{yoshizawa1988new, FINEMET_HITACHI, willard2013nanocrystalline, App_Herzer, dia_lashgari2014composition}, low magnetostriction \cite{yoshizawa1988new, FINEMET_HITACHI, willard2013nanocrystalline, App_Herzer, dia_lashgari2014composition, dia_mattern1995effect}, excellent temperature characteristics \cite{FINEMET_HITACHI}, small aging effects  \cite{FINEMET_HITACHI} and excellent high frequency characteristics \cite{yoshizawa1988new, FINEMET_HITACHI, willard2013nanocrystalline, App_Herzer, dia_lashgari2014composition}, FINEMET alloys have been used for applications such as mobile phones \cite{FINEMET_HITACHI, willard2013nanocrystalline}, noise reduction devices \cite{FINEMET_HITACHI, willard2013nanocrystalline}, computer hard disks \cite{FINEMET_HITACHI, willard2013nanocrystalline} and transformers \cite{yoshizawa1988new, FINEMET_HITACHI, willard2013nanocrystalline, dia_herzer1993nanocrystalline, dia_herzer2013modern, App_Herzer}. Superior soft magnetic properties --in comparison with existing soft magnets at that time,
were achieved by crystallizing Fe$_3$Si nanocrystals ($\alphaα^{''}$-(Fe, Si) phase with D03 structure) from an amorphous matrix.
Optimal properties were achieved for Fe$_3$Si nanocrystals with a mean diameter (radius) between 10-15 nm  (5-7.5 nm) and volume fraction of about 70 \% \cite{yoshizawa1988new, willard2013nanocrystalline, dia_ayers1997model, dia_herzer1993nanocrystalline, dia_van1993nb, dia_herzer1991magnetization, dia_herzer2013modern, dia_herzer2010effect, App_Herzer, dia_herzer2007soft, dia_hono1991atom, dia_lashgari2014composition, dia_clavaguera2002crystallisation, dia_conde1994crystallization, dia_mattern1995effect, dia_hono1999cu}.

Improvements in soft magnetic properties can be achieved by exploring new alloy
compositions and by optimizing the current processing  \cite{yoshizawa1988new, willard2013nanocrystalline, dia_ayers1997model, dia_herzer1993nanocrystalline, dia_van1993nb, dia_herzer1991magnetization, dia_herzer2013modern, dia_herzer2010effect, App_Herzer, dia_herzer2007soft, dia_hono1991atom, dia_lashgari2014composition, dia_clavaguera2002crystallisation, dia_conde1994crystallization, dia_mattern1995effect, dia_hono1999cu}.
One of the challenges faced with this approach is the scarcity of experimental databases for multi-component systems.
Most of the known compositions and associated manufacturing protocols are address the needs of their time.
New experiments and the associated materials characterization can be expensive and time consuming.
Moreover, engineers working with FINEMET alloys often have to deal with multiple, possibly conflicting, objectives in order to design alloys for specific applications.  Hence, it would be beneficial to explore theoretical screening techniques that are based on the physics governing the nucleation and growth, and that can be rapidly and successfully used for testing and screening alloys with compositions not yet studied experimentally.

The CALPHAD approach is one such technique which successfully incorporates advanced models based on several concepts that can explain underlying physics and concomitant phenomena that occur during thermal treatment. In recent years, researchers have used the CALPHAD approach to analyze amorphous phases \cite{palumbo2008thermodynamics} and soft magnets containing amorphous and nanocrystalline phases \cite{takeuchi2014thermodynamic, takeuchi2015thermodynamic, takahashi2017fe} using the commercial software Thermocalc \cite{THERMOCALC}. The reasons behind using the CALPHAD approach, limitations within the databases, and ways to use the databases have been described in these works \cite{takeuchi2014thermodynamic, takeuchi2015thermodynamic, takahashi2017fe}. However, simulations of nucleation and growth of critical phases responsible for soft magnetic properties has not been yet been reported \cite{takeuchi2014thermodynamic, takeuchi2015thermodynamic, takahashi2017fe}.

This motivated us to perform similar investigations in order to  address the problem of simulating nucleation and growth of Fe$_3$Si  nanocrystals (D03 structure)
during isothermal annealing of FINEMET alloys. In this work, we have developed a metastable phase diagram containing the Fe$_3$Si phase, and a precipitation model capable of simulating nucleation and growth of Fe$_3$Si nanocrystals during annealing at several reported temperatures \cite{yoshizawa1988new, willard2013nanocrystalline, dia_ayers1997model, dia_herzer1993nanocrystalline, dia_van1993nb, dia_herzer1991magnetization, dia_herzer2013modern, dia_herzer2010effect, App_Herzer, dia_herzer2007soft, dia_hono1991atom, dia_lashgari2014composition, dia_clavaguera2002crystallisation, dia_conde1994crystallization, dia_mattern1995effect, dia_hono1999cu}.
In the model, we were able to crystallize Fe$_3$Si nanocrystals with a desired mean radius (5-7.5 nm) and volume fraction (70\%) \cite{yoshizawa1988new, willard2013nanocrystalline, dia_ayers1997model, dia_herzer1993nanocrystalline, dia_van1993nb, dia_herzer1991magnetization, dia_herzer2013modern, dia_herzer2010effect, App_Herzer, dia_herzer2007soft, dia_hono1991atom, dia_lashgari2014composition, dia_clavaguera2002crystallisation, dia_conde1994crystallization, dia_mattern1995effect, dia_hono1999cu} by performing isothermal annealing at a set of annealing temperatures for one hour holding time.

\section{METHODS}\label{methods}

\begin{figure}[ht]
\centering
\includegraphics[width = 7.4cm]{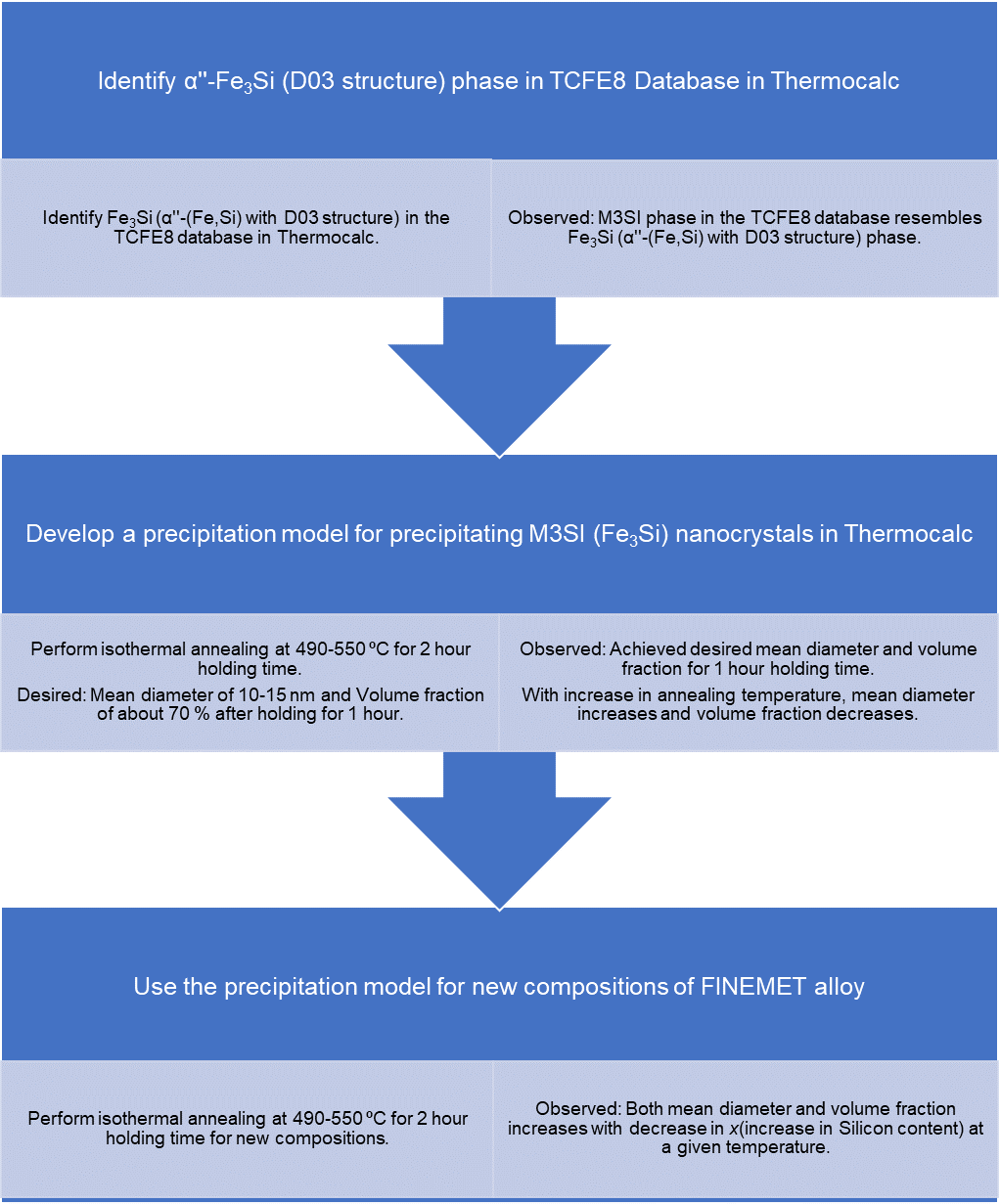}
\caption{Schematic flowchart outlining the present work on modeling the precipitation kinetics of Fe$_3$Si nanocrystals in FINEMET.}
\label{Paper_1_Flowchart}
\end{figure}

Figure \ref{Paper_1_Flowchart} shows the flowchart describing the main steps of this work.

We used the Thermocalc \cite{THERMOCALC} software for calculating the
equilibrium and metastable phase diagrams, and the TC-PRISMA module in Thermocalc to develop a precipitation model capable of simulating nucleation and growth of the Fe$_3$Si.  We used the thermodynamic database TCFE8 \cite{THERMOCALC_TCFE8} for equilibrium calculations, and the mobility database MOBFE3 \cite{THERMOCALC_TCFE8_MOBFE3} for simulating the precipitation kinetics of Fe$_3$Si nanocrystals. We paid particular attention to ensure that we simulated the nucleation and growth of the
same phase (i.e., Fe$_3$Si) that we observed when deriving the metastable phase diagram for the Fe-Si system.
Subsequent simulations of nucleation and growth of Fe$_3$Si nanocrystals during isothermal annealing between 490 \degree C and 550 \degree C were performed through the developed precipitation model for 1-2 h holding time. The precipitation model was further used for studying nucleation and growth of Fe$_3$Si nanocrystals during  annealing between 490 \degree C and 550 \degree C for compositions in the vicinity of the given composition,  Fe$_{72.89+x}$Si$_{16.21-x}$B$_{6.90}$Nb$_{3}$Cu$_{1}$ (where $x$ = $\pm$ 3 atomic \%).

\subsection{Phase Diagrams}

There is no Fe$_3$Si phase in the TCFE8 database, so this exact phase cannot appear in the phase diagrams.
However, based on stoichiometry, we surmise that that M3Si phase in TCFE8 could be the Fe$_3$Si phase. This phase, however,
does not appear in the equilibrium phase diagram of our alloy system.
We performed calculations of Gibbs energy-Composition diagram at 540 \degree C to identify the metastable phases, and found that
the M3Si phase is indeed among them. This lead us to perform phase diagram calculations so as to preferentially stabilize the M3Si phase.
From our plotted metastable phase diagram and reported works on Fe-Si system \cite{fesi_kubaschewski1993phase, fesi_kiyokane2017b2, fesi_matsumura2001concurrent, fesi_silveyra2011structural}, we observed that the M3Si region in TCFE8 database coincides with the previously reported region where Fe$_3$Si occurs.
Hence, in this work we used M3SI phase to simulate the nucleation and growth of Fe$_3$Si nanocrystals.

\subsection{Time-Temperature-Transformation Diagram of FINEMET}\label{Section_TTT_diagram_FINEMET}

The FINEMET composition used here is Fe$_{72.89}$Si$_{16.21}$B$_{6.90}$Nb$_{3}$Cu$_{1}$ in atomic \%, or Fe$_{82.35}$Si$_{9.21}$B$_{1.51}$Nb$_{5.64}$Cu$_{1.29}$ in weight \%.

\begin{figure}[H]
\centering
\includegraphics[width = 8.4 cm]{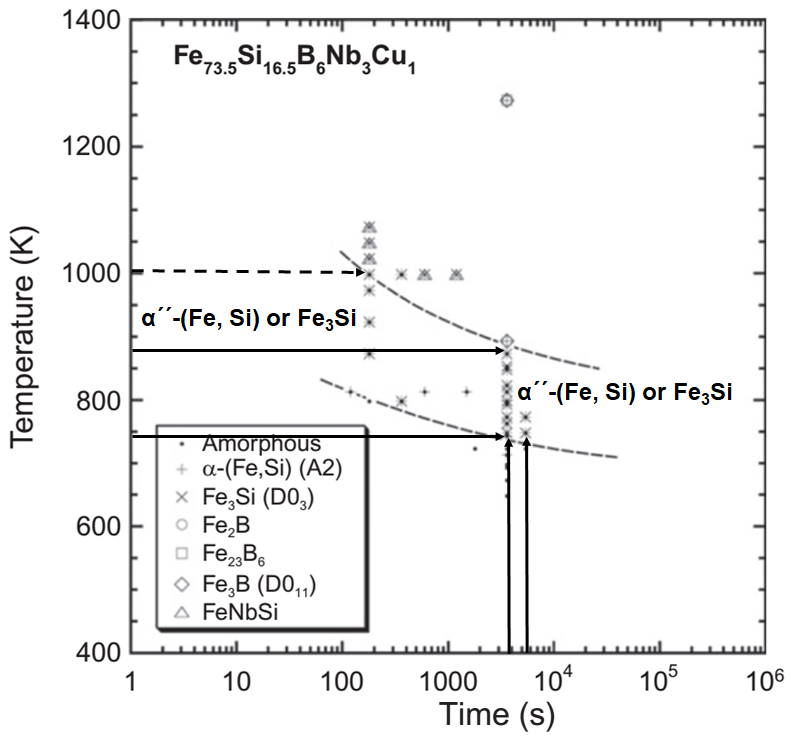}
\caption{TTT diagram for FINEMET of composition Fe$_{73.5}$Si$_{16.5}$B$_{6}$Nb$_{3}$Cu$_{1}$ (Adapted from Willard and Daniil,~\cite{willard2013nanocrystalline} with permission from Elsevier)}
\label{TTT_Diagram_FINEMET}
\end{figure}

Figure \ref{TTT_Diagram_FINEMET} shows the time-temperature-transformation (TTT) diagram of a very similar alloy, Fe$_{73.5}$Si$_{16.5}$B$_{6}$Nb$_{3}$Cu$_{1}$ \cite{willard2013nanocrystalline}.
From Figure \ref{TTT_Diagram_FINEMET}, we note that it is possible to crystallize only Fe$_3$Si for temperatures between 773 K (500 \degree C ) and 833 K (560 \degree C) for holding times between 1.0 and 1.5 h.
Additionally, we note that Fe$_3$Si nanocrystals can be crystallized by annealing for about 200 s holding time at temperatures between about  833 K (560 \degree C) and 1000 K (727 \degree C). But isothermal annealing at this temperature for 1 h holding time leads to precipitation of intermetallic Fe-B phases that need to be avoided as they are detrimental for soft magnetic properties \cite{ttt_yoshizawa1991magnetic, ttt_herzer1997nanocrystalline}. To ensure that no detrimental intermetallic Fe-B phase will be precipitated, in this work we focused on performing isothermal annealing at  490 \degree C, 500 \degree C, 510 \degree C, 520 \degree C, 530 \degree C, 540 \degree C and 550 \degree C for 2 h holding time.

%%(Adapted from Ref.~\onlinecite{willard2013nanocrystalline} with permission from Elsevier)

\subsection{Precipitation model development using the Thermocalc TC-PRISMA module}\label{methods_ppt_model_development}

Significant amount of work has been reported on modeling of amorphous phases using the
CALPHAD approach \citep{palumbo2008thermodynamics}.
There are often specific assumptions \cite{palumbo2008thermodynamics} that are to be followed so as to obtain precipitation of nanocrystals from an
amorphous precursor. For modeling of the amorphous phase, one approach is to treat it as an under-cooled liquid below
the glass transition temperature \cite{palumbo2008thermodynamics}.
As mentioned in Sec.~\ref{introduction}, soft magnets containing amorphous phases have been successfully
analyzed using Thermocalc\cite{takeuchi2014thermodynamic, takeuchi2015thermodynamic, takahashi2017fe}, but precipitation kinetics
of nucleation and growth of critical phases has not been reported.
This motivated us  to develop a model capable of simulating the nucleation and growth of Fe$_3$Si nanocrytals.

In the CALPHAD community, several models are extensively used for simulating precipitation kinetics of various phases in metallic and amorphous systems \cite{palumbo2008thermodynamics, COMPUTHERM}. The precipitation module TC-PRISMA  \cite{THERMOCALC_TCPRISMA} is based on Langer-Schwartz theory and it uses the Kampmann-Wagner Numerical (KWN) approach \cite{kampmann1984decomposition, wagner1991homogeneous} to simulate the simultaneously occurring phenomena of nucleation, growth, and coarsening. The Langer and Schwartz theory \cite{THERMOCALC_TCPRISMA, COMPUTHERM, langer1980kinetics} is a fast-acting model  used for simulation of particle number density and mean size of the precipitated phase. The KWN method \cite{kampmann1984decomposition, wagner1991homogeneous} is an extension of the Langer-Schwartz approach and its modified form \cite{COMPUTHERM}. KWN is also used for prediction of particle size distribution (PSD) over the full course of precipitation \cite{THERMOCALC_TCPRISMA, COMPUTHERM}. Furthermore, this module can analyze the size distribution of precipitates over the complete thermal treatment protocol applied for multicomponent systems and multiphase alloys \cite{COMPUTHERM}.

For the modeling of the amorphous phase, we considered the LIQUID phase in TCFE8 as the matrix.
As mentioned, the precipitate phase is M3SI in TCFE8 database \citep{THERMOCALC_TCFE8}.
During development of the precipitation model, most of the parameters were set at default values
in TC-PRISMA, \cite{THERMOCALC_TCPRISMA} as our purpose was to develop the precipitation model with least number of user-defined inputs.
User-defined inputs that we varied in order to develop the precipitation model are the
molar volume, the nucleation sites, and the mobility enhancement prefactor.
From experiments, we found the density of FINEMET ribbons as 8.35 g/cm$^{3}$.
Based on this density value, the molar volume is 5.913 $\times$ $10^{-6}$ m$^{3}$/mol.
which we used for the LIQUID phase. The Nucleation sites were set to occur in the bulk (in TC-PRISMA, a user can set nucleation sites in
bulk, at grain boundaries, or at dislocations).
The mobility enhancement prefactor was set by trial and error at 2 $\times$ $10^{-10}$.
%In the Thermocalc manual, it is mentioned that this value takes intergranular diffusion into consideration. In the MOBFE3 database, it is mentioned that for modelling the LIQUID phase, the default value to consider diffusion is $ 10^{-9} $. So, we worked around this value to calibrate our model. Additionally, Thermocalc community motivates users to perform calculations in spite of limitations mentioned in their manual \cite{THERMOCALC_TCFE8_MOBFE3}.

\section{RESULTS AND DISCUSSION}\label{Results}
In this section, we start by determining the region in the Fe-Si phase diagram where the Fe$_3$Si phase is stable (Sec.~\ref{Sec_Fe_Si_Diagram}).
This is followed by results obtained after performing isothermal annealing at various temperatures through the precipitation model (Sec.~\ref{res_Simulating_Fe3Si}). Finally, we used this model for simulating nucleation and growth of Fe$_3$Si nanocrystals by isothermal annealing at various temperatures for
new compositions that are in the vicinity of the given composition of FINEMET alloy (Sec.~\ref{res_Vary_Comp}).

\subsection{Equilibrium and Metastable Phase Diagrams}\label{Sec_Fe_Si_Diagram}
The phases of the Fe-Si system that are present in the TCFE8 database are listed in Table \ref{tab_Phases_Fe_Si_diagram},
 along with their composition and lattice occupancy.

\begin{table*}[ht]
%\tiny
  \centering
  \caption{Composition and lattice occupancy of various phases shown in the Fe-Si phase diagram (in lattice occupancy, VA means Vacancy)}
\begin{tabular}{llllll}
%%%\toprule
\hline
\hline
    Phase & \multicolumn{4}{c}{Atomic \%} & Lattice Occupancy \\
    \hline
			     & & Fe & Si &   \\
\hline
%    %    \midrule
     B2\_BCC & & 0.915 & 0.085 & & $(Fe,Si)_{0.5}$ $(Fe,Si)_{0.5}$ $(VA)_{3}$ \\
%    %    \midrule
     BCC\_A2 & & 0.915 & 0.085 & & $(Fe,Si)_{1}$ $(VA)_{3}$ \\
%    %    \midrule
     M3SI & & 0.750 & 0.250 & & $(Fe)_{3}$ $(Si)_{1}$ \\
%   %    \midrule
     FCC\_A1 & & 0.915 & 0.085 & & $(Fe,Si)_{1}$ $(VA)_{1}$ \\
%    %    \midrule
     HCP\_A3 & & 0.935 & 0.065 & & $(Fe,Si)_{1}$ $(VA)_{0.5}$ \\
%    %    \midrule
     FE2SI & & 0.667 & 0.333 & & $(Fe)_{0.67}$ $(Si)_{0.33}$ \\
%    %    \midrule
     M5SI3 & & 0.625 & 0.375 & & $(Fe)_{0.62}$ $(Si)_{0.38}$ \\
%    %    \midrule
     MSI & & 0.500 & 0.500 & & $(Fe)_{0.5}$ $(Si)_{0.5}$ \\
%    %    \midrule
     DIAMOND\_FCC\_A4 & & 0.000 & 1.000 & & $(Si)_{1}$ \\
     \hline
%    %    \bottomrule
 \end{tabular} \label{tab_Phases_Fe_Si_diagram}%
\end{table*}

\begin{figure*}[ht]
\centering
\includegraphics[width=12cm]{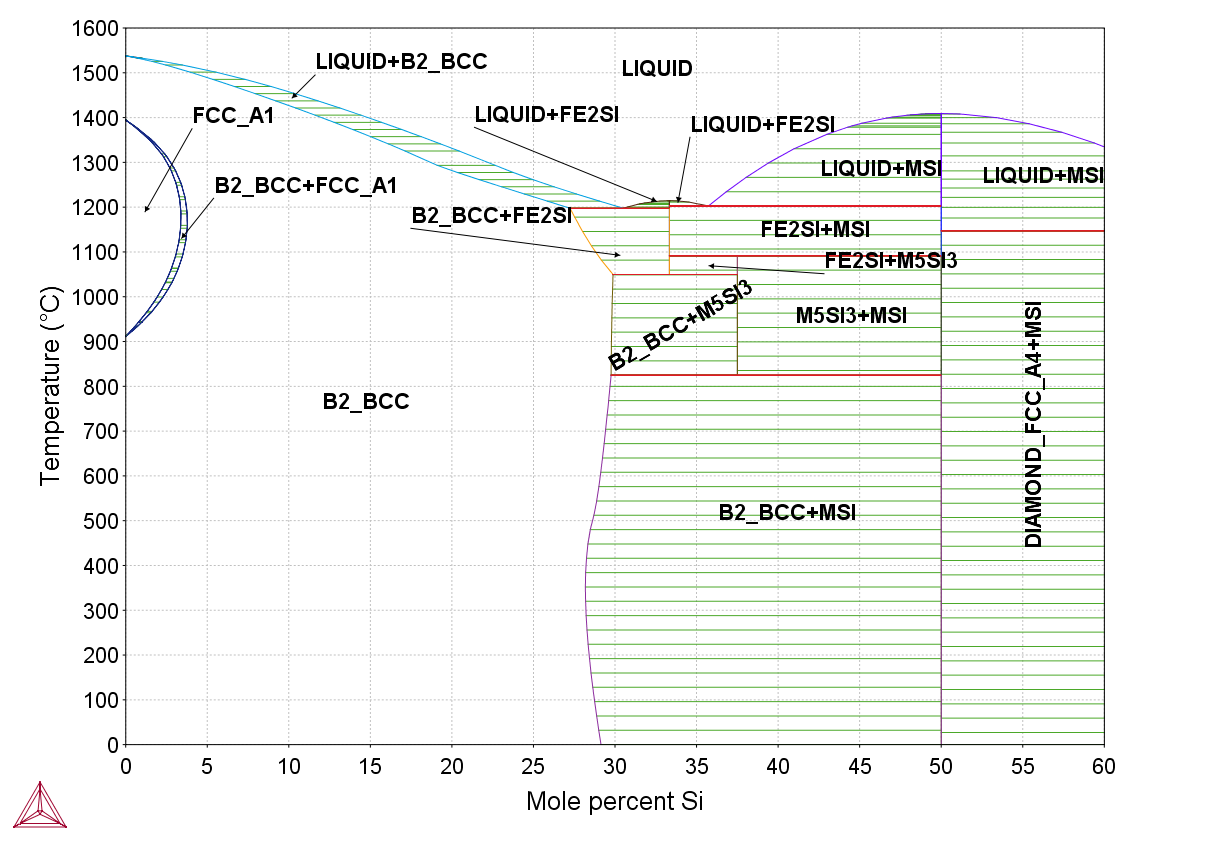}
\caption{The equilibrium Fe-Si phase diagram showing the B2\_BCC phase in a large range of temperatures and Si compositions.}
\label{Fe_SI_Phase_B2_BCC}
\end{figure*}

\begin{figure*}[ht]
\centering
\includegraphics[width=12cm]{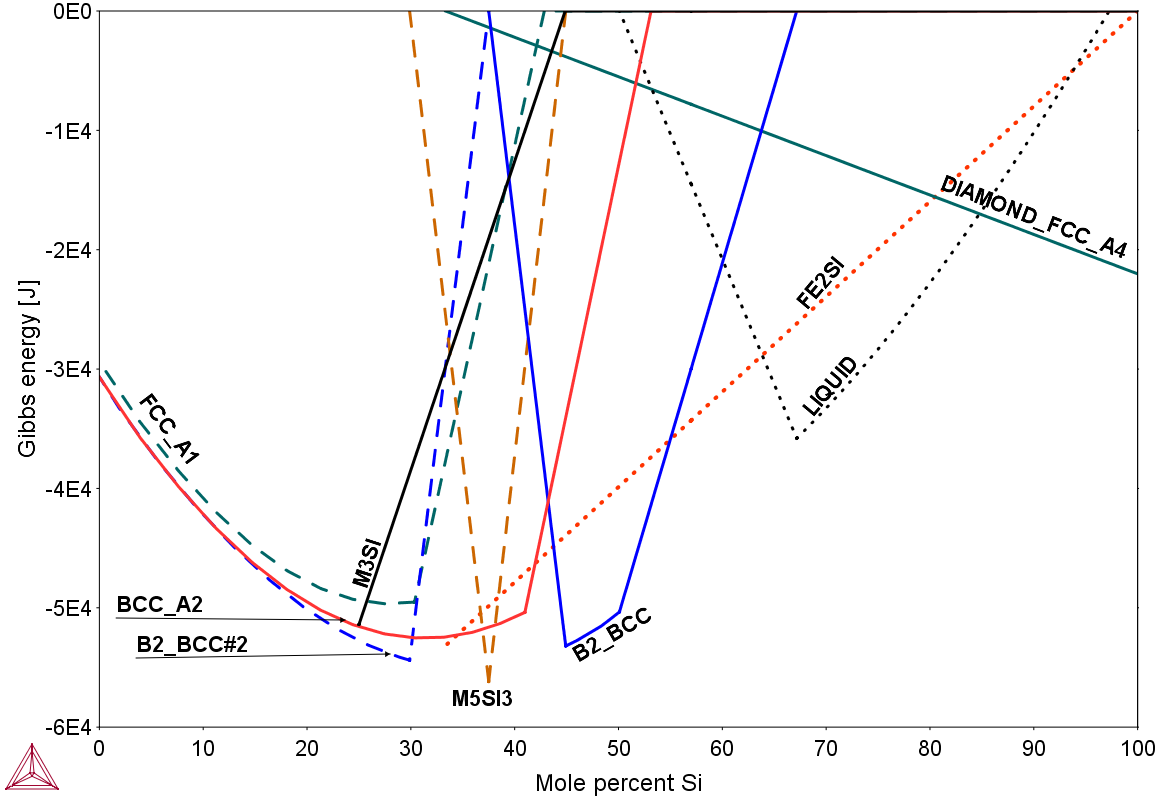}
\caption{Gibbs Energy-Composition diagram ($G$-$x$) at 540 \degree C showing M3SI, BCC\_A2, and  B2\_BCC phase, along with some phases listed in Table~\ref{tab_Phases_Fe_Si_diagram}.}
\label{Final_Fe_SI_Phase_G-X}
\end{figure*}

BCC Fe-Si exists in three forms, namely A2 ($\alphaα$-(Fe, Si)), B2 ($\alphaα^{'}$-(Fe, Si)) and D03 ($\alphaα^{''}$-(Fe, Si)).
First, we performed equilibrium calculations for the Fe-Si system and plotted the equilibrium Fe-Si phase diagram (Figure \ref{Fe_SI_Phase_B2_BCC}).
We note that  B2\_BCC is stable up to 30 mole \% Si at $\SI{1050}{\degreeCelsius}$ as a single phase.
From $\SI{900}{\degreeCelsius}$ to about $\SI{1400}{\degreeCelsius}$, FCC\_A1 is stable as a single phase for up to 3 mole \% Si.
There is a narrow region between $\SI{900}{\degreeCelsius}$ to about $\SI{1400}{\degreeCelsius}$ in which both  B2\_BCC and FCC\_A1 exist.
From the lattice occupancy in Table \ref{tab_Phases_Fe_Si_diagram}, we note that M3SI resembles Fe$_3$Si phase, but M3SI phase does not show in the equilibrium phase diagram (Figure \ref{Fe_SI_Phase_B2_BCC}). Similarly, BCC\_A2 does not appear in the diagram (Figure \ref{Fe_SI_Phase_B2_BCC}).
This means that both BCC\_A2 and M3SI could be metastable phases.
We therefore performed calculations at 540 \degree C to plot the Gibbs Energy-Composition (or G-X) diagram in order to identify the metastable phases present at 540 \degree C (Figure \ref{Final_Fe_SI_Phase_G-X}).
%Gibbs Energy-Composition (or G-X) diagram will be helpful to identify the phases that needs to be removed in order to preferentially stabilize M3SI phase and plot it on the Fe-Si metastable phase diagram.

\begin{figure*}[ht]
\centering
\includegraphics[width=12cm]{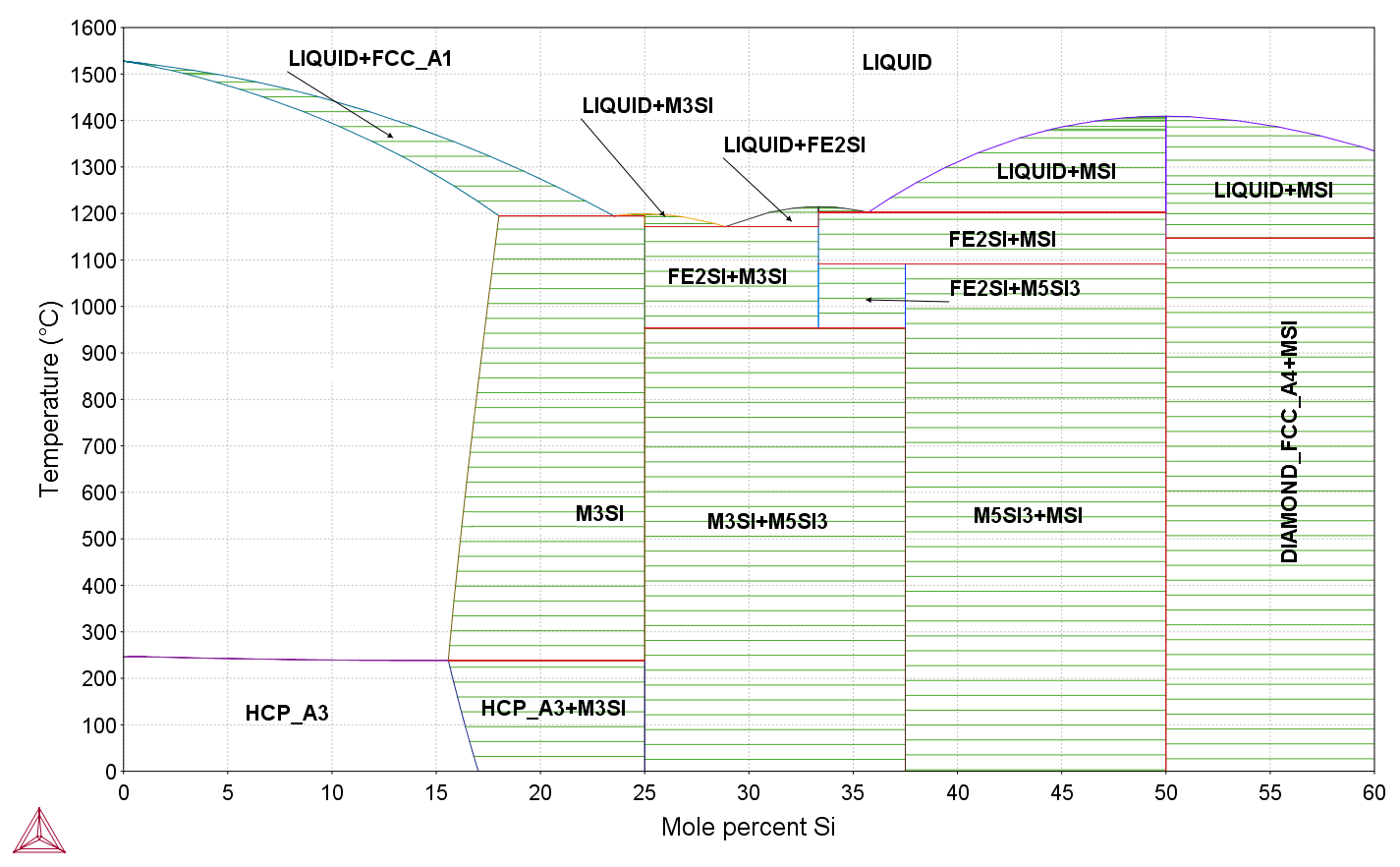}
\caption{Metastable Fe-Si phase diagram showing the M3SI phase, obtained by suppressing  the B2\_BCC and BCC\_A2 phases from the equilibrium phase diagram.)}
\label{Fe_SI_Phase_M3SI}
\end{figure*}

In Figure \ref{Final_Fe_SI_Phase_G-X}, we observe that at 540 \degree C, M3SI phase has a higher  Gibbs free energy when compared to the BCC\_A2 and B2\_BCC phases. Thus, M3SI is a metastable phase relative to BCC\_A2 and B2\_BCC, which is why it does not appear on the Fe-Si equilibrium phase diagram (Figure~ \ref{Fe_SI_Phase_B2_BCC}). Before we proceed, it is important to identify the temperature-composition region in which M3SI can appear in the Fe-Si phase diagram.
In Thermocalc, all computed phase diagrams are equilibrium ones, hence they will not contain metastable phases.
In order to find the conditions for which M3Si can exist, we circumvent this by suppressing one or more phases from equilibrium calculation.
Based on the G-x plots (Figure \ref{Final_Fe_SI_Phase_G-X}), we remove the B2\_BCC and BCC\_A2 phases from the phase diagram calculations, and re-evaluate the remaining
competing phases. The result is the metastable phase diagram in Figure \ref{Fe_SI_Phase_M3SI}, which shows that the M3Si phase
exists in the region between about 15 to 25 mole \% Si; also,  the M3SI phase co-exists with the M5SI3 phase from 25 - 37 mole \% Si.

%In Figure \ref{Fe_SI_Phase_B2_BCC}, we can observe that FCC\_A1 is confined to a small region, whereas in Figure \ref{Fe_SI_Phase_M3SI}, we can observe FCC\_A1 phase extending to the region out of the bounds it was confined in Figure \ref{Fe_SI_Phase_B2_BCC}. This is due to the fact that while performing equilibrium calculations to plot M3SI (Fe$_3$Si) phase (Figure \ref{Fe_SI_Phase_M3SI}), we removed B2\_BCC and BCC\_A2 as these were the stable phases in Gibbs energy-Composition diagram calculations (Figure \ref{Final_Fe_SI_Phase_G-X}). We didn't remove FCC\_A1 while performing calculations as M3SI (Fe$_3$Si) phase cuts through FCC\_A1 in Figure \ref{Final_Fe_SI_Phase_G-X}, thus we we didn't remove it from the equilibrium calculations at 540 \degree C. Hence in absence of B2\_BCC and BCC\_A2, FCC\_A1 phase becomes stable and extends beyond the bounds it was confined in Figure \ref{Fe_SI_Phase_B2_BCC} and thus we can also observe a mixture of M3SI (Fe$_3$Si) and FCC\_A1 phase in Figure \ref{Fe_SI_Phase_M3SI}.

On comparing Figure \ref{Fe_SI_Phase_M3SI} with works studying both the Fe-Si binary system and the FINEMET alloys \cite{fesi_kubaschewski1993phase, fesi_kiyokane2017b2, fesi_matsumura2001concurrent, fesi_silveyra2011structural}, we notice that the region of occurrence of the M3SI phase coincides with the region in which the Fe$_3$Si phase has been observed. This confirms that the M3SI  phase can be used to simulate nucleation and growth of Fe$_3$Si nanocrystals from the amorphous phase (Sec. \ref{methods_ppt_model_development}).

%The main objective of Figure \ref{Fe_SI_Phase_M3SI} is to identify the region in which M3SI ( Fe$_3$Si) phase is stable as it is the desired phase in this work. From TTT diagram in Figure \ref{TTT_Diagram_FINEMET}, we already know that M3SI (Fe$_3$Si) phase can be precipitated as a single phase by isothermal annealing at $\SI{540}{\degreeCelsius}$ (813 K) for 1 hour and 30 minutes holding time and no inter-metallic Fe-B phase can be precipitated even for prolonged holding time. Thus, Figure \ref{Fe_SI_Phase_M3SI} will provide with additional information that will prove to be helpful during designing heat treatment schedule for precipitating M3SI ( Fe$_3$Si) nanocrystals in the desired size range with optimum volume fraction required for superior magnetic properties. From Figure \ref{Fe_SI_Phase_M3SI} and  the reported works on Fe-Si system  \cite{fesi_kubaschewski1993phase, fesi_kiyokane2017b2, fesi_matsumura2001concurrent, fesi_silveyra2011structural}, we have successfully identified the region of occurrence of Fe$_3$Si ($\alphaα^{''}$-(Fe, Si) D03) phase on the Fe-Si phase diagram and confirmed that it resembles with M3SI (Fe$_3$Si) phase reported in TCFE8 database \citep{THERMOCALC_TCFE8}. Additionally, we reported the composition range in which M3SI (Fe$_3$Si) phase is stable. In Sec.~\ref{res_Simulating_Fe3Si}, we have reported the results obtained after using this model for performing isothermal annealing for a set of annealing temperatures for 2 hour holding time for a given composition of FINEMET alloy mentioned in Sec.~\ref{methods_ppt_model_development}.

\subsection{Simulating Nucleation and Growth of M3SI (Fe$_3$Si) in TC-PRISMA}\label{res_Simulating_Fe3Si}

Based on guidelines in Sec.~\ref{methods_ppt_model_development}, we developed a precipitation model
capable of simulating nucleation and growth of M3SI (Fe$_3$Si) nanocrystals in the TC-PRISMA  module \cite{THERMOCALC_TCPRISMA} of Thermocalc \cite{THERMOCALC}.
The TTT diagram (Figure \ref{TTT_Diagram_FINEMET}) shows that it is possible to precipitate only Fe$_3$Si  for isothermal annealing at 540 \degree C (813 K) for 1.5 h
holding time, with no inter-metallic Fe-B phase appearing even for prolonged holding times.
In order to examine the effects of chaging the temperature, we used the developed model to perform isothermal annealing at 490, 500, 510, 520, 530, 540 and 550 \degree C
for precipitating M3SI (Fe$_3$Si) nanocrystals for 2 h, at the nominal composition of Fe$_{72.89}$Si$_{16.21}$B$_{6.90}$Nb$_{3}$Cu$_{1}$ (atomic \%).

\begin{figure}[ht]
\centering
\includegraphics[width = 8.4 cm]{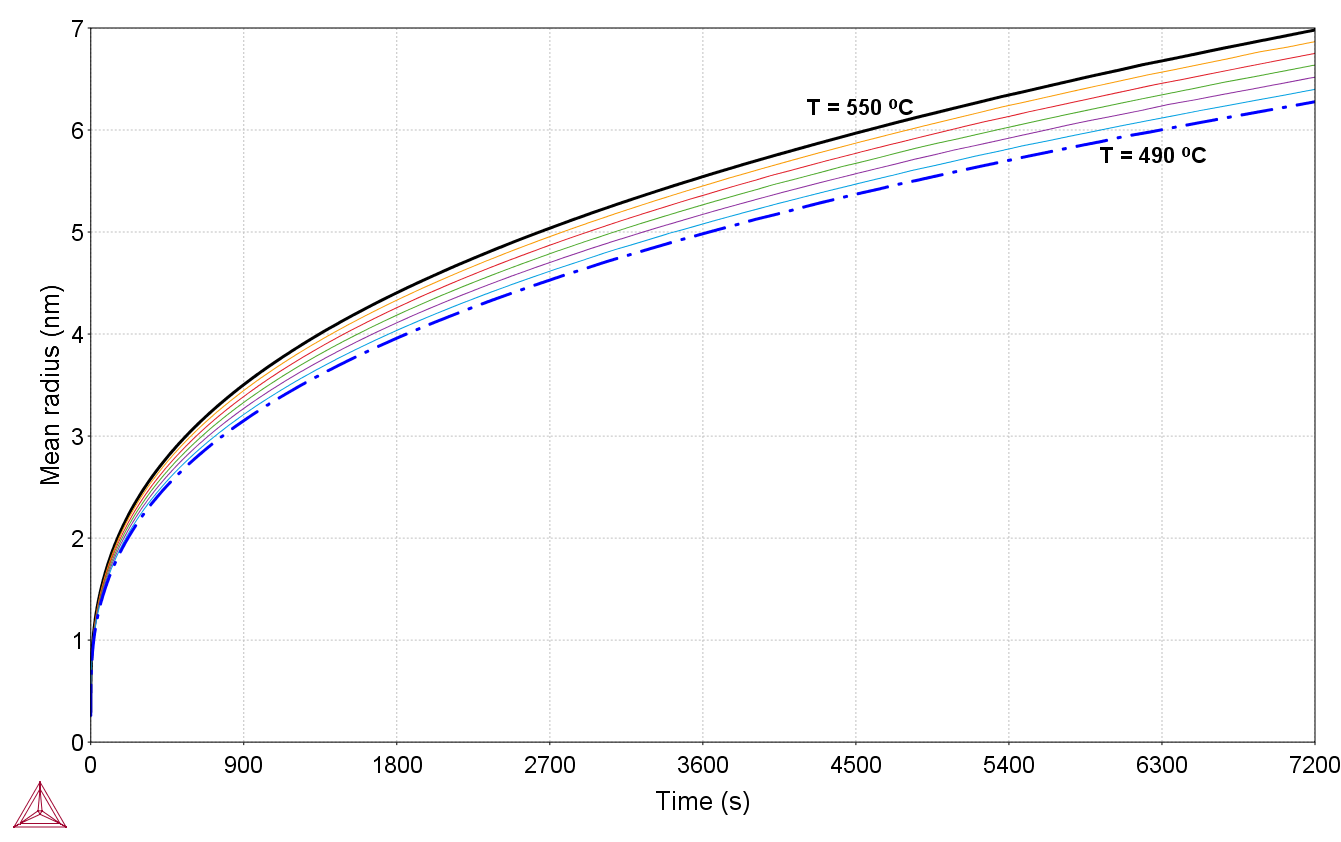}
\caption{Mean radius of Fe$_3$Si nanocrystals vs. time during isothermal annealing at various annealing temperatures.}
\label{Mean_rad_Time_Initial}
\end{figure}

\begin{figure}[ht]
\centering
\includegraphics[width = 8.4 cm]{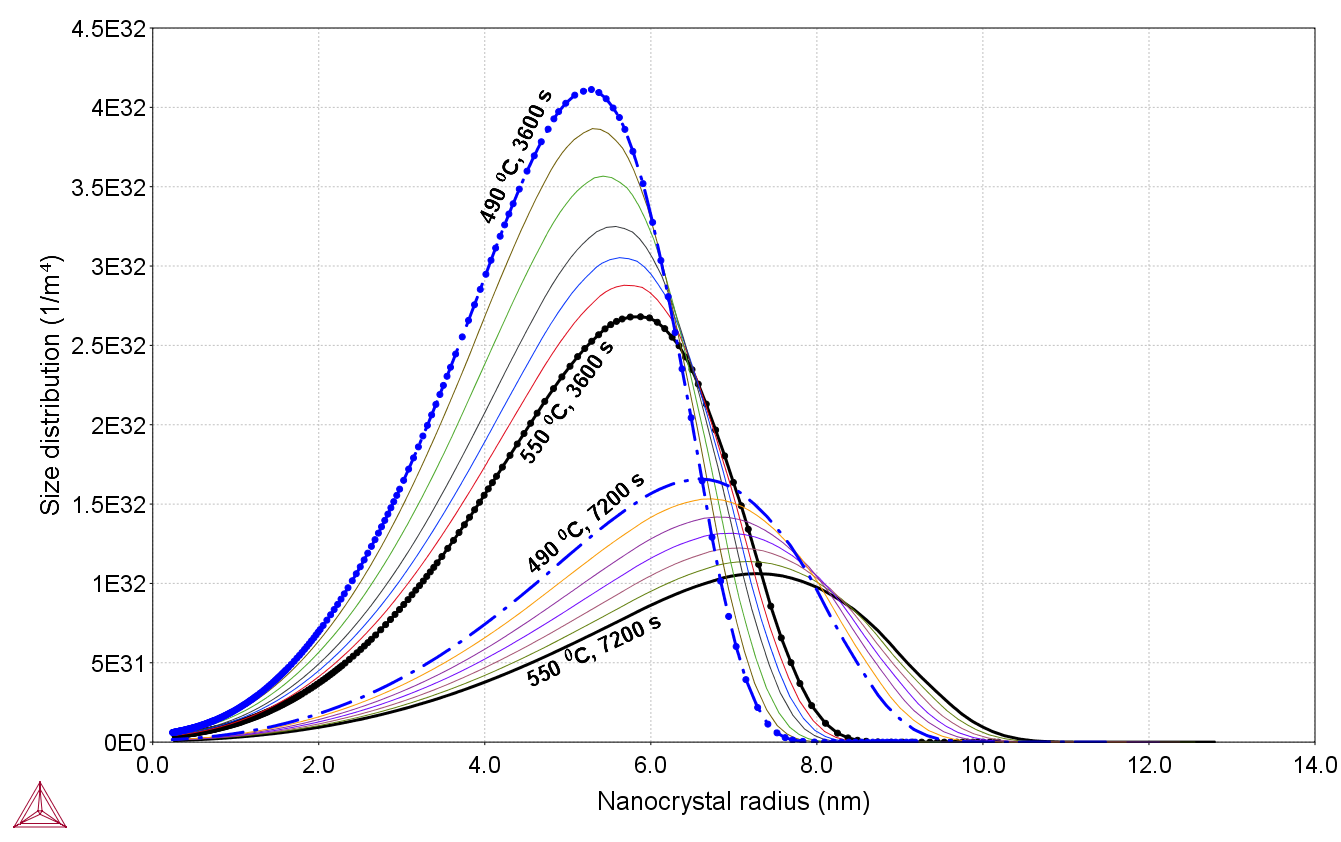}
\caption{Size distribution of Fe$_3$Si nanocrystals obtained after isothermal annealing at various annealing temperatures for 1 h and 7200 h holding time.}
\label{Size_distribution_Initial}
\end{figure}

\subparagraph{Mean Radius vs Time}\label{Mean_radius_Results}
Figure \ref{Mean_rad_Time_Initial} shows the variation of mean radius of M3SI (Fe$_3$Si) nanocrystals as a function of holding time
for various annealing temperatures. At 1 h holding time, we were able to achieve a mean radius above 5 nm but below 6 nm (Table  \ref{tab_Mean_diameter_Time}),
which is in the desired range \cite{yoshizawa1988new, willard2013nanocrystalline, dia_ayers1997model, dia_herzer1993nanocrystalline, dia_van1993nb, dia_herzer1991magnetization, dia_herzer2013modern, dia_herzer2010effect, App_Herzer, dia_herzer2007soft, dia_hono1991atom, dia_lashgari2014composition, dia_clavaguera2002crystallisation, dia_conde1994crystallization, dia_mattern1995effect, dia_hono1999cu}. Another important observation
is that mean radius increases with the annealing temperature Figure \ref{Mean_rad_Time_Initial}, which is also apparent in
 a plot of the size distribution plot (Figure \ref{Size_distribution_Initial}).

%Mean radius was used to calculate mean diameter of  M3SI (Fe$_3$Si) nanocrystals obtained after isothermal annealing at various annealing temperatures at 3600 s, 5400 s and 7200 s holding time and has been tabulated in Table \ref{tab_Mean_diameter_Time}. In Table \ref{tab_Mean_diameter_Time}, it can be seen that the mean diameter of  M3SI (Fe$_3$Si) nanocrystals is in the desired range of 10-15 nm \cite{willard2013nanocrystalline, dia_ayers1997model,nuc_size_sharma2015competition, dia_herzer1993nanocrystalline, dia_van1993nb, dia_hampel1995structure, dia_herzer1991magnetization, dia_herzer2013modern, dia_herzer2010effect, dia_herzer2007soft, dia_hono1991atom, dia_lashgari2014composition, dia_clavaguera2002crystallisation, dia_conde1994crystallization, dia_mattern1995effect, dia_hono1999cu}.

\begin{table*}[ht]
  \centering
  \caption{Mean radius of Fe$_3$Si nanocrystals during isothermal annealing at various annealing temperatures for different holding times}

\begin{tabular}{lllllllllllllll}
%\toprule
\hline
\hline
     Temp. ($^{\circ}$C) & \multicolumn{5}{c}{Mean radius (nm)} & \multicolumn{5}{c}{Volume fraction} & \multicolumn{3}{c}{Silicon (mole \%)}\\
     \hline
			    & 1.0 h & 1.5 h s & 2.0 h & & & 1.0 h & 1.5 h s & 2.0 h & & & 1.0 h & 1.5 h  & 2.0 h \\
\hline
    %    \midrule
    490 & 5.0 & 5.75 & 6.3 & & & 0.7060 & 0.7064 & 0.7066 & & & 1.809 & 1.787 & 1.771 \\
    %    \midrule
    500 & 5.1 & 5.85 & 6.4 & & & 0.7038 & 0.7041 & 0.7045 & & & 1.947 & 1.927 & 1.904 \\
    %    \midrule
    510 & 5.2 & 5.95 & 6.5 & & & 0.7015 & 0.7020 & 0.7022 & & & 2.085 & 2.059 & 2.046 \\
    %    \midrule
    520 & 5.3 & 6.05 & 6.65 & & & 0.6989 & 0.6995 & 0.6998 & & & 2.244 & 2.210 & 2.190 \\
    %    \midrule
    530 & 5.4 & 6.15 & 6.75 & & & 0.6962 & 0.6969 & 0.6973 & & & 2.404 & 2.365 & 2.340 \\
    %    \midrule
    540 & 5.5 & 6.25 & 6.85 & & & 0.6936 & 0.6944 & 0.6947 & & & 2.557 & 2.513 & 2.496 \\
    %    \midru
    550 & 5.6 & 6.35 & 7.0 & & & 0.6909 & 0.6913 & 0.6919 & & & 2.716 & 2.689 & 2.660 \\
    %    \bottomrule
\hline
%%    \bottomrule
\end{tabular} \label{tab_Mean_diameter_Time}%
\end{table*}%

%\begin{table}[ht]
 % \centering
 % \caption{Mean Diameter of M3SI (Fe$_3$Si) nanocrystals during isothermal annealing at various annealing temperatures for 3600, 5400 and 7200 seconds holding time}

%\begin{tabular}{lllll}
%\toprule
%\hline
%\hline
 %    Temp. ($^{\circ}$C) & \multicolumn{3}{c}{Mean Diameter (nm)} \\
%     \hline
%			    & 3600 s & 5400 s & 7200 s  \\
%\hline
    %    \midrule
%    490 & 9.997 & 11.479 & 12.555 \\
%    %    \midrule
  %  500 & 10.235 & 11.686 & 12.797 \\
 %   %    \midrule
 %   510 & 10.424 & 11.889 & 13.034 \\
    %    \midrule
 %   520 & 10.577 & 12.113 & 13.271 \\
%    %    \midrule
%    530 & 10.732 & 12.262 & 13.502 \\
    %    \midrule
%    540 & 10.967 & 12.547 & 13.733 \\
    %    \midrule
%    550 & 11.153 & 12.747 & 13.964 \\
    %    \bottomrule
%\hline
%%    \bottomrule
%\end{tabular} \label{tab_Mean_diameter_Time}%
%\end{table}%

\subparagraph{Size Distribution}\label{Result_size_dist}
Figure \ref{Size_distribution_Initial} shows the size distribution of M3SI (Fe$_3$Si) nanocrystals for various annealing temperatures for 1 h and 2 h holding times. For 1 h  holding time, the maximum mean radius is around 8 nm, while for 2 h holding time the mean radius is about 10 nm.
In the latter case, the mean size is out of the the desired range.
The area under the size distribution plot for any of the curves provides the number density of
M3SI (Fe$_3$Si) nanocrystals/$ m^{3} $ with nanocrystal sizes within the range shown in Figure \ref{Size_distribution_Initial} for a particular isothermal annealing temperature and holding time \cite{psd_zender2010particle}. The number density obtained for the precipitated M3SI (Fe$_3$Si) nanocrystals is
consistent with those reported in literature \cite{dia_clavaguera2002crystallisation, dia_hono1999cu}.
Thus, for the current annealing conditions, we were able to obtain optimum mean radius by isothermal annealing at all the annealing temperatures
under consideration for 1 h time.
We can now proceed to estimating the volume fraction of M3SI (Fe$_3$Si) obtained after isothermal annealing through our model.

\begin{figure}[ht]
\centering
\includegraphics[width = 8.4 cm]{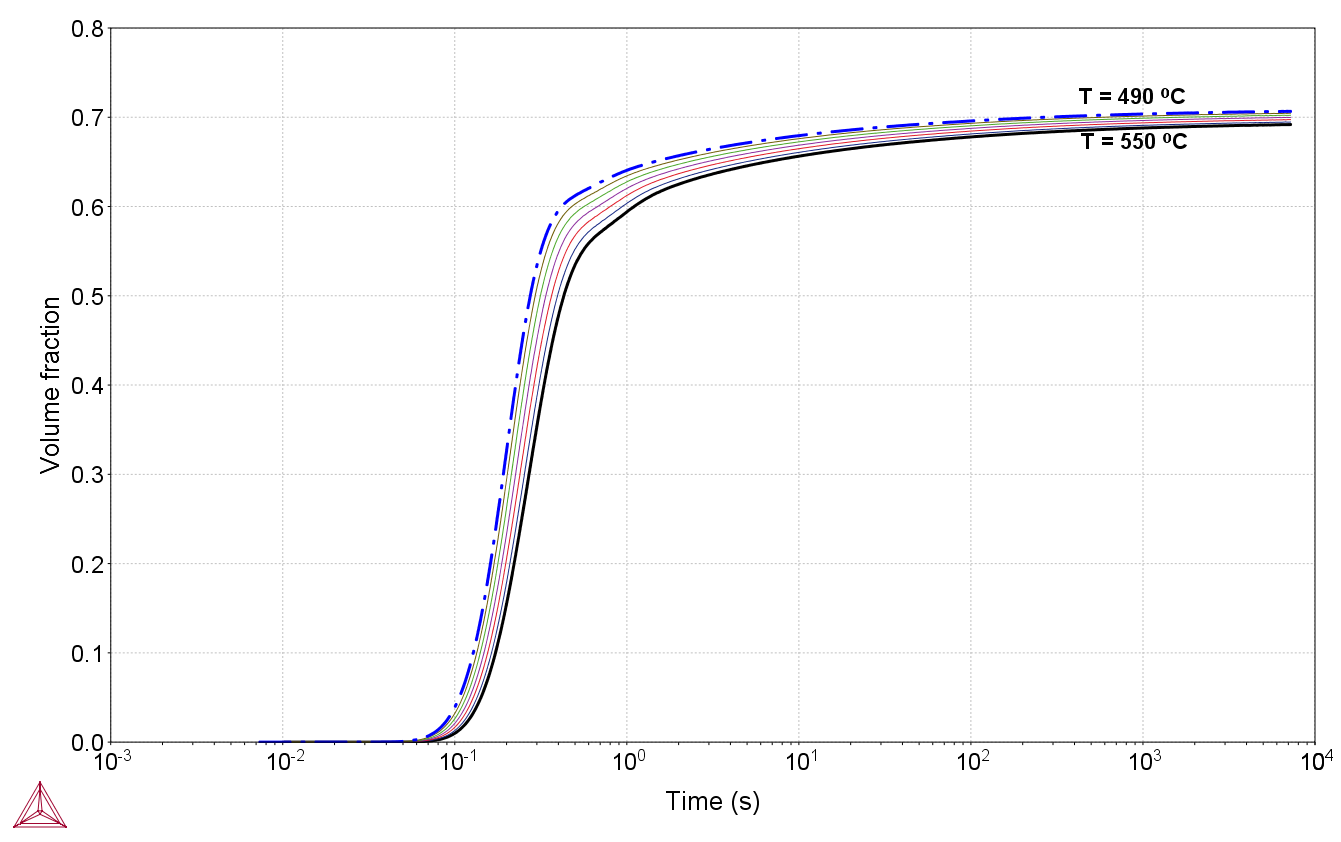}
\caption{Volume fraction of M3SI (Fe$_3$Si) nanocrystals vs time during isothermal annealing at various annealing temperatures.}
\label{Volume_fraction_Time_Initial}
\end{figure}

\subparagraph{ Volume Fraction vs. Time}\label{Res_Vol_Frac}
In Sec.~\ref{introduction}, we mentioned that the  target nanocrystalline volume fraction is about 70 \% \citep{willard2013nanocrystalline, dia_lashgari2014composition, dia_clavaguera2002crystallisation, dia_mattern1995effect}. Figure \ref{Volume_fraction_Time_Initial} shows the variation of volume fraction as a function of time for during isothermal annealing at various annealing temperatures for up to 1 h holding time. We were able to achieve the desired volume fraction
of 70 \% for all temperatures under consideration after 1 h annealing.
The volume fraction obtained after 1 h, 1.5 h, and 2 h has been tabulated in Table \ref{tab_Mean_diameter_Time}.
%Variation of volume is extremely small when compared to volume fraction obtained after annealing at the annealing temperatures mentioned in Table \ref{tab_Mean_diameter_Time}, that's why the volume fraction is reported in Table \ref{tab_Mean_diameter_Time} for up to 4 places after the decimal.
The volume fraction of precipitated M3SI (Fe$_3$Si) nanocrystals decreases with increase in annealing temperatures for the same holding time.

Additional information regarding the nucleation rate (Appendix \ref{a_nuc_rate}), number density (Appendix \ref{a_number_density}), and driving force (Appendix \ref{a_driving_force}) for crystallization of M3SI (Fe$_3$Si) nanocrystals has been included in the Appendix \ref{appendix}.
This information can be used to support the volume fraction results shown here. The observed nucleation rate and number density is
in accordance with reported experimental work on FINEMET alloys \cite{dia_clavaguera2002crystallisation, dia_hono1999cu}.

%%\begin{table}[ht]
%%  \centering
%%  \caption{Volume fraction of Fe$_3$Si nanocrystals during isothermal annealing at various annealing temperatures for 3600, 5400 and 7200 seconds holding time}

%%\begin{tabular}{lllll}
%%%\toprule
%%\hline
%%\hline
%%    Temp. ($^{\circ}$C) & \multicolumn{3}{c}{Volume Fraction} \\
%%    \hline
%%			    & 3600 s & 5400 s & 7200 s  \\
%%\hline
%%    %    \midrule
%%    490 & 0.7060 & 0.7064 & 0.7066 \\
%%    %    \midrule
%%    500 & 0.7038 & 0.7041 & 0.7045 \\
%%    %    \midrule
%%    510 & 0.7015 & 0.7020 & 0.7022 \\
%%    %    \midrule
%%    520 & 0.6989 & 0.6995 & 0.6998 \\
%%    %    \midrule
%%    530 & 0.6962 & 0.6969 & 0.6973 \\
%%    %    \midrule
%%    540 & 0.6936 & 0.6944 & 0.6947 \\
%%    %    \midrule
%%    550 & 0.6909 & 0.6913 & 0.6919 \\
%%    %    \bottomrule
%%\hline
%%%%    \bottomrule
%%\end{tabular} \label{tab_Vol_Fraction}%
%%\end{table}%

\begin{figure}[ht]
\centering
\includegraphics[width = 8.4 cm]{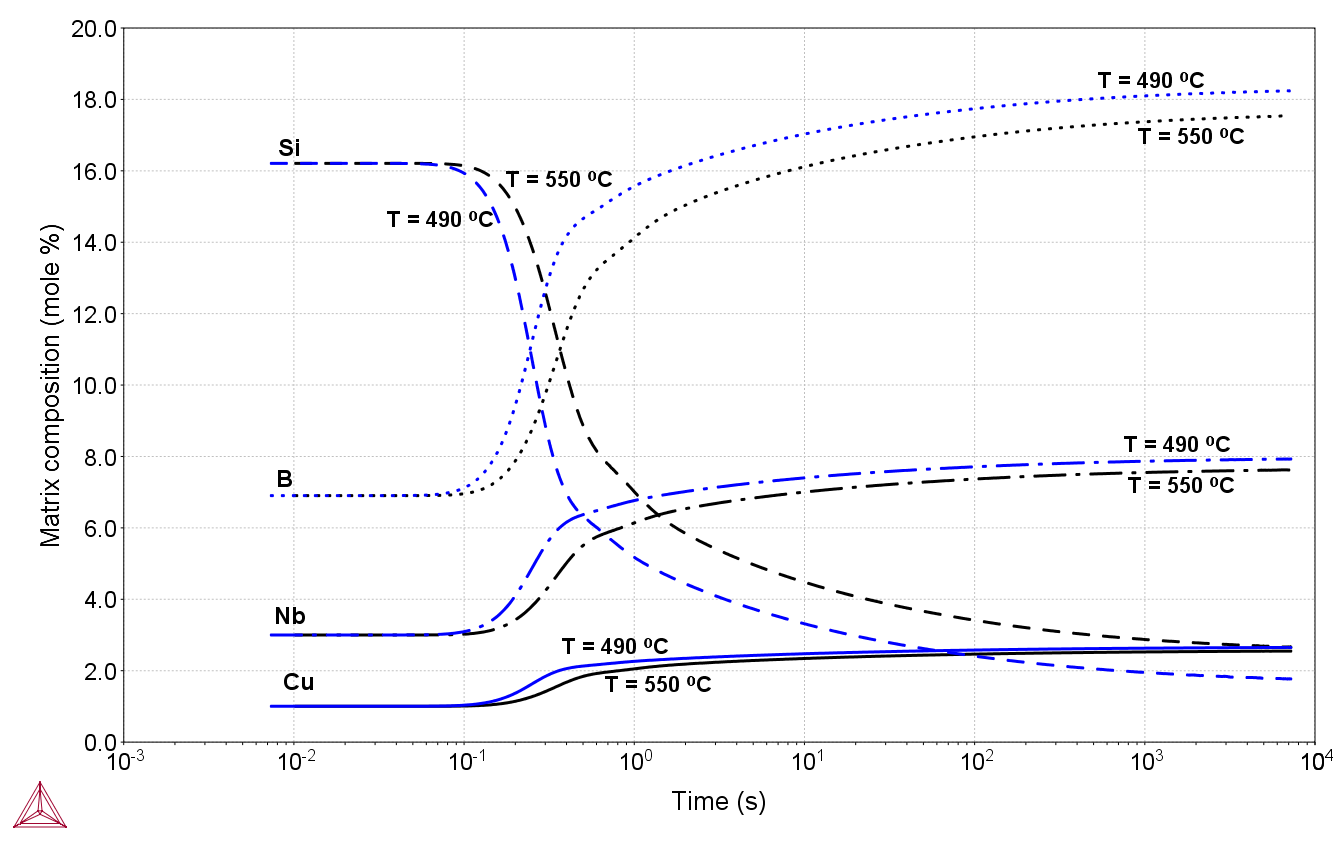}
\caption{Matrix Composition vs time during isothermal annealing at various annealing temperatures.}
\label{Matrix_composition_Time_Initial}
\end{figure}

\subparagraph{Matrix Composition vs. Time}\label{res_Matrix_comp}
Figure \ref{Matrix_composition_Time_Initial} shows the variation of matrix composition for annealing at 490 and 550 \degree C. We can observe that there is a significant difference in matrix composition of various elements for the two temperatures over the course of isothermal annealing for 2 h. The Si content is significantly lower for annealing at 490 \degree C as compared to annealing at 550 \degree C. This can be understood from the fact that the volume fraction of the M3SI (Fe$_3$Si) phase is higher at 490 \degree C when compared to volume fraction at 550 \degree C (Figure \ref{Volume_fraction_Time_Initial} and Table \ref{tab_Mean_diameter_Time} in Sec. ~\ref{Res_Vol_Frac}). Thus, more silicon went into the formation of  M3SI (Fe$_3$Si) nanocrystals and hence, less Si remained in the matrix at 490 \degree C. The matrix composition of Si obtained after isothermal annealing for 1, 1.5, and 2 h holding time for various annealing temperatures is listed in Table \ref{tab_Mean_diameter_Time}.
These values for niobium, boron and copper have been tabulated and are included in Appendix \ref{a_matrix_comp}, Table \ref{tab_Matrix_Nb}.

%%\begin{table}[ht]
%%  \centering
%%  \caption{Matrix Composition of Silicon during isothermal annealing at two annealing temperatures for 3600, 5400 and 7200 seconds holding time}

%%\begin{tabular}{lllll}
%%%\toprule
%%\hline
%%\hline
%%    Temp. ($^{\circ}$C) & \multicolumn{3}{c}{Silicon (mole \%)} \\
%%    \hline
%%			    & 3600 s & 5400 s & 7200 s  \\
%%\hline
%%    %    \midrule
%%    490 & 1.809 & 1.787 & 1.771 \\
%%    %    \midrule
%%    500 & 1.947 & 1.927 & 1.904 \\
%%    %    \midrule
%%    510 & 2.085 & 2.059 & 2.046 \\
%%    %    \midrule
%%    520 & 2.244 & 2.210 & 2.190 \\
 %%   %    \midrule
%%    530 & 2.404 & 2.365 & 2.340 \\
%%    %    \midrule
%%    540 & 2.557 & 2.513 & 2.496 \\
%%    %    \midrule
%%    550 & 2.716 & 2.689 & 2.660 \\
%%    %    \bottomrule
%%\hline
%%\end{tabular} \label{tab_Matrix_Si}%
%%\end{table}%

\subsection{Using the precipitation model for Fe$_3$Si nanocrystals from matrices of different compositions}\label{res_Vary_Comp}
\begin{figure}[ht]
\centering
\includegraphics[width = 8.4 cm]{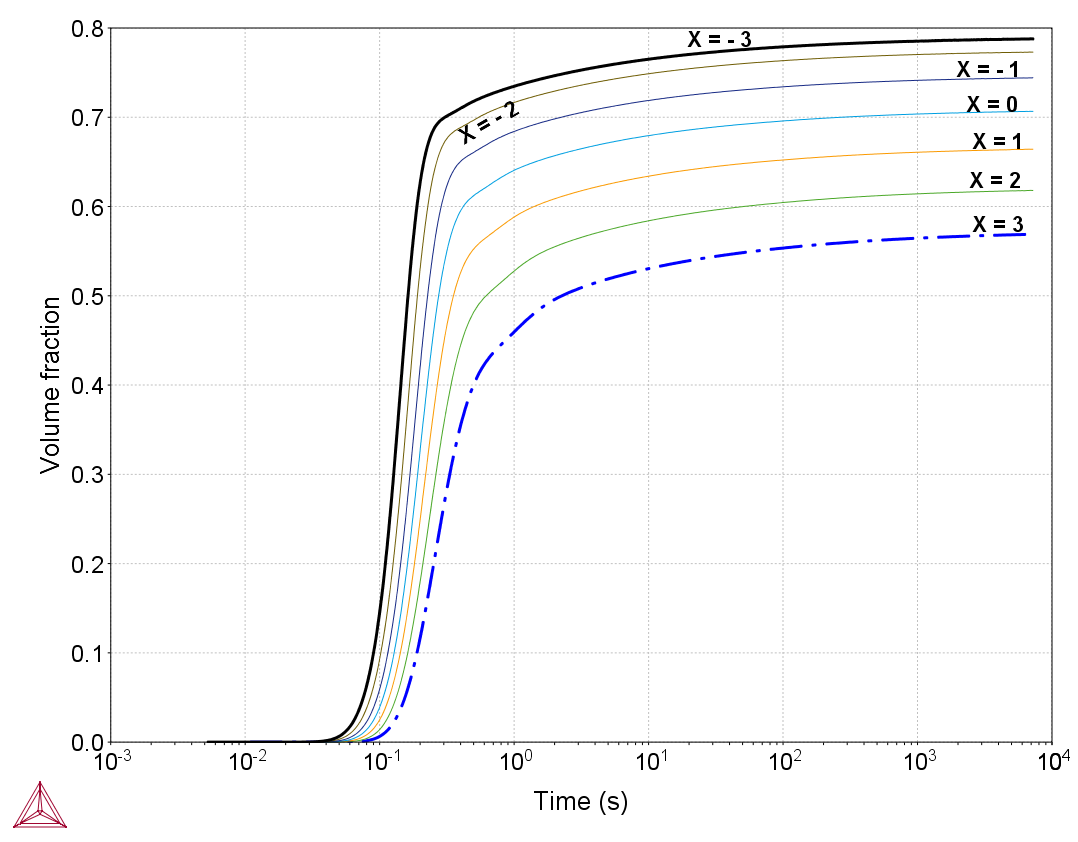}
\caption{Volume Fraction (of M3SI (Fe$_3$Si) nanocrystals) vs Time during isothermal annealing at $\SI{490}{\degreeCelsius}$ for various compositions.}
\label{Volume_fraction_Time_vary_X}
\end{figure}

In Sec.~\ref{res_Simulating_Fe3Si}, we have shown the scope of application of our
proposed CALPHAD model as an effective predictive tool and used it for simulating nucleation and growth of M3SI (Fe$_3$Si) nanocrystals from
an amorphous precursor.
We can also apply this model to  FINEMET alloys of different compositions. To this end, we
 altered the variable bounds of Fe and Si by $ \pm $ 3 atomic \%, and performed isothermal annealing at the temperatures mentioned in Sec.~\ref{res_Simulating_Fe3Si},
 for 2 h holding time.
 The new compositions can be written as Fe$_{72.89+x}$Si$_{16.21-x}$B$_{6.90}$Nb$_{3}$Cu$_{1}$ (where $x$ = $\pm$ 3 atomic \% or $x$ = [3, 2, 1, 0, -1, -2, -3]).
The following figures show the variation of mean radius (Figure \ref{Mean_rad_Time_vary_X}) and volume fraction (Figure \ref{Volume_fraction_Time_vary_X}) during isothermal annealing at $\SI{490}{\degreeCelsius}$ for 2 h holding time  for these compositions.

\subparagraph{Mean Radius for varying composition}
Figure \ref{Mean_rad_Time_vary_X} shows the  mean radius of M3SI (Fe$_3$Si) nanocrystals during
isothermal annealing at 490 $^{\circ}$C for a 2 h holding time
for various $x$ values.
As $x$ increases (Fe increases and Si decreases), mean radius decreases for isothermal annealing at 490 \degree C
displaying a saturation behavior as the mean radius for $x$ = 2 and $x$ = 3 are very close to one another.
For $x$ $ < $ 0,
the mean radius obtained after annealing at 490 \degree C is less than 5 nm, i.e. outside the desired range.
Therefore, for $x$ $ < $ 0, one has to increase the annealing time above 1 h for to obtain a mean radius above 5 nm.

\begin{figure}[ht]
\centering
\includegraphics[width = 8.4 cm]{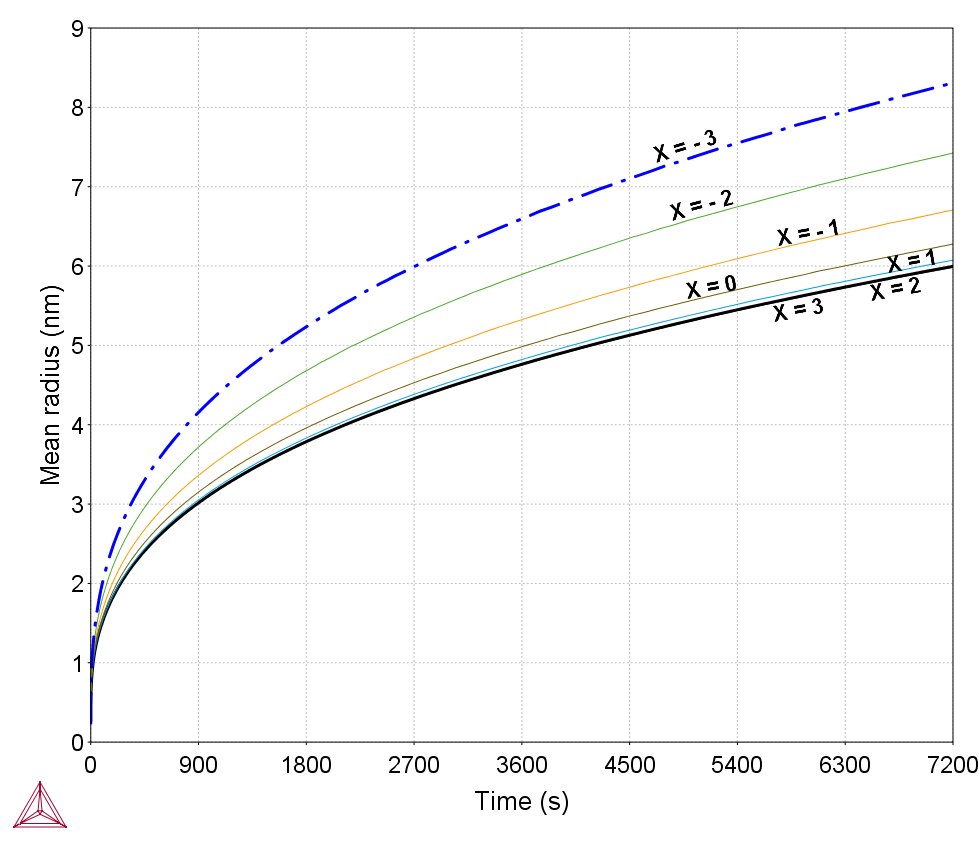}
\caption{Mean radius of M3SI (Fe$_3$Si) nanocrystals vs time during isothermal annealing at $\SI{490}{\degreeCelsius}$ for various compositions.}
\label{Mean_rad_Time_vary_X}
\end{figure}

\subparagraph{Volume fraction for varying composition}
Figure \ref{Volume_fraction_Time_vary_X} shows the variation of volume fraction of M3SI (Fe$_3$Si) nanocrystals during isothermal annealing at 490 $^{\circ}$C for a 2 h holding time for various composition of FINEMET obtained by varying $x$.
As $x$ increases (Fe increases and Si decreases), the volume fraction decreases for isothermal annealing at 490 \degree C. Additionally, for $x$ $ < $ 0, the volume fraction obtained after isothermal annealing at 490 \degree C is less than the desired  70 \%, even for prolonged holding time. At $x$ = 3, the volume fraction is as low as 55\%. Still this volume fraction is within the range of volume fraction reported in other studies on FINEMET alloys \cite{willard2013nanocrystalline, dia_lashgari2014composition, dia_clavaguera2002crystallisation, dia_mattern1995effect}.

From Figures \ref{Mean_rad_Time_vary_X} and \ref{Volume_fraction_Time_vary_X}, we note that  both mean radius and volume fraction decrease with increasing $x$.
This trend was also observed for isothermal annealing at $\SI{500}{\degreeCelsius}$, $\SI{510}{\degreeCelsius}$, $\SI{520}{\degreeCelsius}$, $\SI{530}{\degreeCelsius}$, $\SI{540}{\degreeCelsius}$ and $\SI{550}{\degreeCelsius}$. Thus, it is important to focus on Fe-Si content to achieve the desired mean radius and volume fraction for optimum properties.

\section{CONCLUSIONS}\label{Conclusions}
In this work, we laid the guidelines for developing a precipitation model within Thermocalc for simulating nucleation and growth of  Fe$_3$Si nanocrystals from an amorphous precursor of soft magnetic FINEMET alloy. Fe$_3$Si phase was identified as the M3Si in the TCFE8 database
by plotting metastable phases on the Fe-Si phase diagram and comparing the metastable M3SI phase in TCFE8 database with various studies on Fe-Si system that had
reported the Fe$_3$Si (D03) phase.
Thereafter, we used this M3SI phase and developed a precipitation model for studying the nucleation and growth of magnetic M3SI(Fe$_3$Si) nanocrystals during isothermal annealing for a set of annealing temperatures (490-550 \degree C) for up to 2 h holding time.
During isothermal annealing, the mean radius increases with increase in annealing temperature while volume fraction decreases with increase in temperature.

Results obtained from precipitation model during isothermal annealing at various annealing temperatures (490-550 \degree C) correlate
well with the observations reported in the literature regarding mean radius, size range and volume fraction. Thus, we proceeded with using this precipitation model for new compositions that are in the vicinity of the nominal FINEMET composition.
Isothermal annealing was performed on these new compositions for the same set of annealing temperatures (490 \degree C - 550 \degree C), for 2 h holding time.
We found that with decrease in Silicon content, both mean radius and volume fraction increases for all the annealing temperatures under consideration.

In conclusion, we developed a robust precipitation model under the framework of CALPHAD approach that is capable of simulating nucleation and growth of M3SI or Fe$_3$Si nanocrystals from an amorphous precursor for a class of FINEMET alloy by isothermal annealing at a set of annealing temperatures (490 \degree C - 550 \degree C) for 2 hour holding time. Subject to careful parametrization, this model can be used to study the precipitation of crystalline phases
for other alloys with large number of atomic species, or even for the same FINEMET alloys but for different crystals (e.g., copper nanoparticles that are believed to
act as nucleation sites for the D03 phase).

\section{ACKNOWLEDGMENT}
Authors acknowledge the support of National Science Foundation through Grant No. DMREF-1629026.

%\begin{appendices}
\section{APPENDIX}\label{appendix}

%\section{Nanocrystal Precipitation}
\subsection{Results}\label{a_results}
Results reported in this section support the findings reported in Sec.~\ref{res_Simulating_Fe3Si}.

\subsubsection{Nucleation Rate vs. Time} \label{a_nuc_rate}
\begin{figure}[ht]
\centering
\includegraphics[width = 8.4 cm]{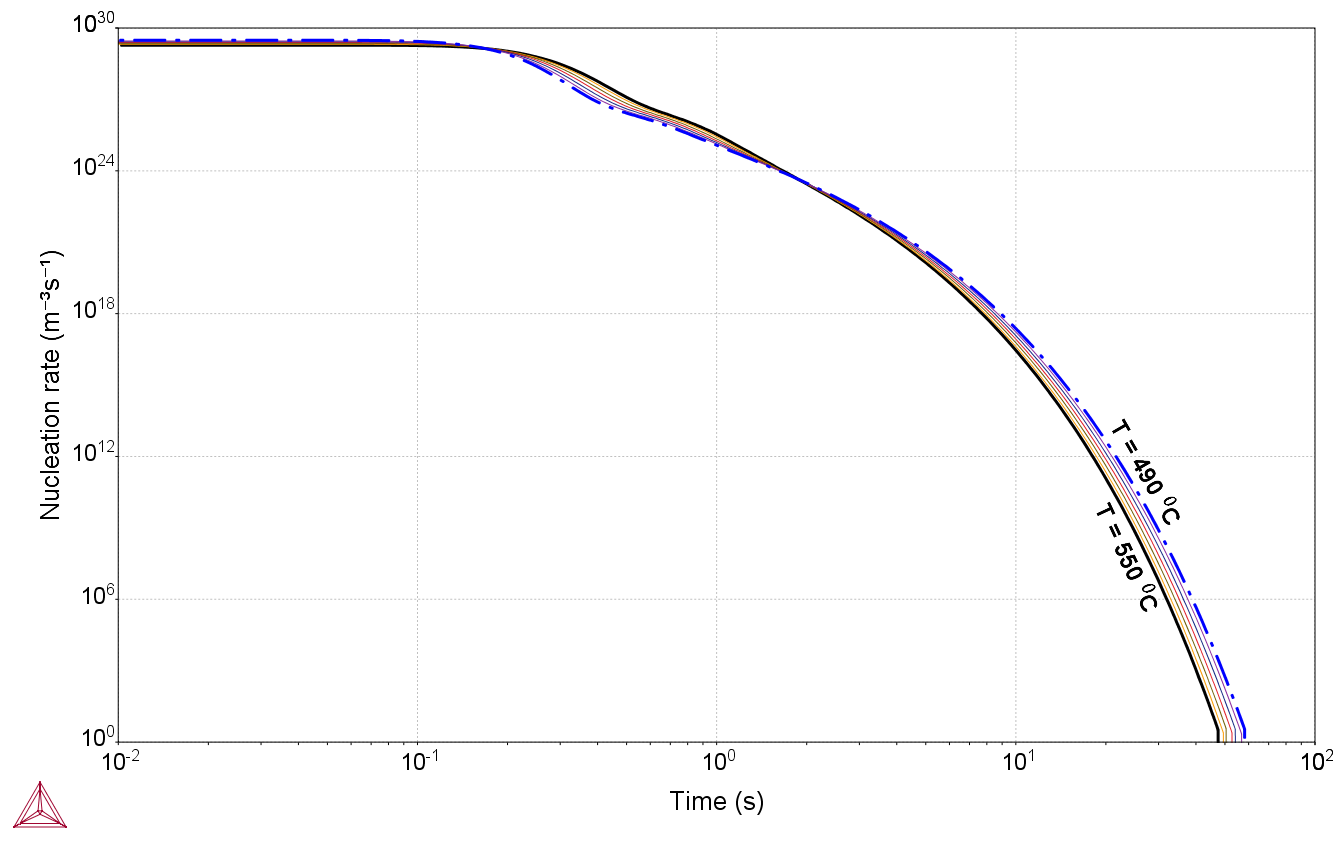}
\caption{Nucleation Rate (of M3SI (Fe$_3$Si)) vs Time during isothermal annealing at various annealing temperatures.}
\label{Nucleation_rate_Time_Initial}
\end{figure}

Figure \ref{Nucleation_rate_Time_Initial} shows the plot for variation of nucleation rate of M3SI (Fe$_3$Si) nanocrystals during isothermal annealing at various annealing temperatures (490-550 \degree C) for 2 hour holding time. It can be observed that nucleation ends after about 1 minute. Additionally, nucleation rates decreases with increase in temperature, where the nucleation rate decreases by 6 orders of magnitude in a few seconds. This plot can be helpful in understanding the volume fraction plot (Figure \ref{Volume_fraction_Time_Initial}) and Table \ref{tab_Mean_diameter_Time} reported in Sec.~\ref{res_Simulating_Fe3Si}, where we observed that volume fraction decreases with increase in temperature. After one minute, radius will increase by growth mechanism of preexisting nucleus and hence, less number of nucleus at higher temperature can be responsible for comparatively less volume fraction obtained when annealing is performed at elevated temperatures. Observed nucleation rate is in accordance with reported literature on FINEMET alloys \cite{dia_clavaguera2002crystallisation, dia_hono1999cu}.

\subsubsection{Number Density vs. Time} \label{a_number_density}
\begin{figure}[ht]
\centering
\includegraphics[width = 8.4 cm]{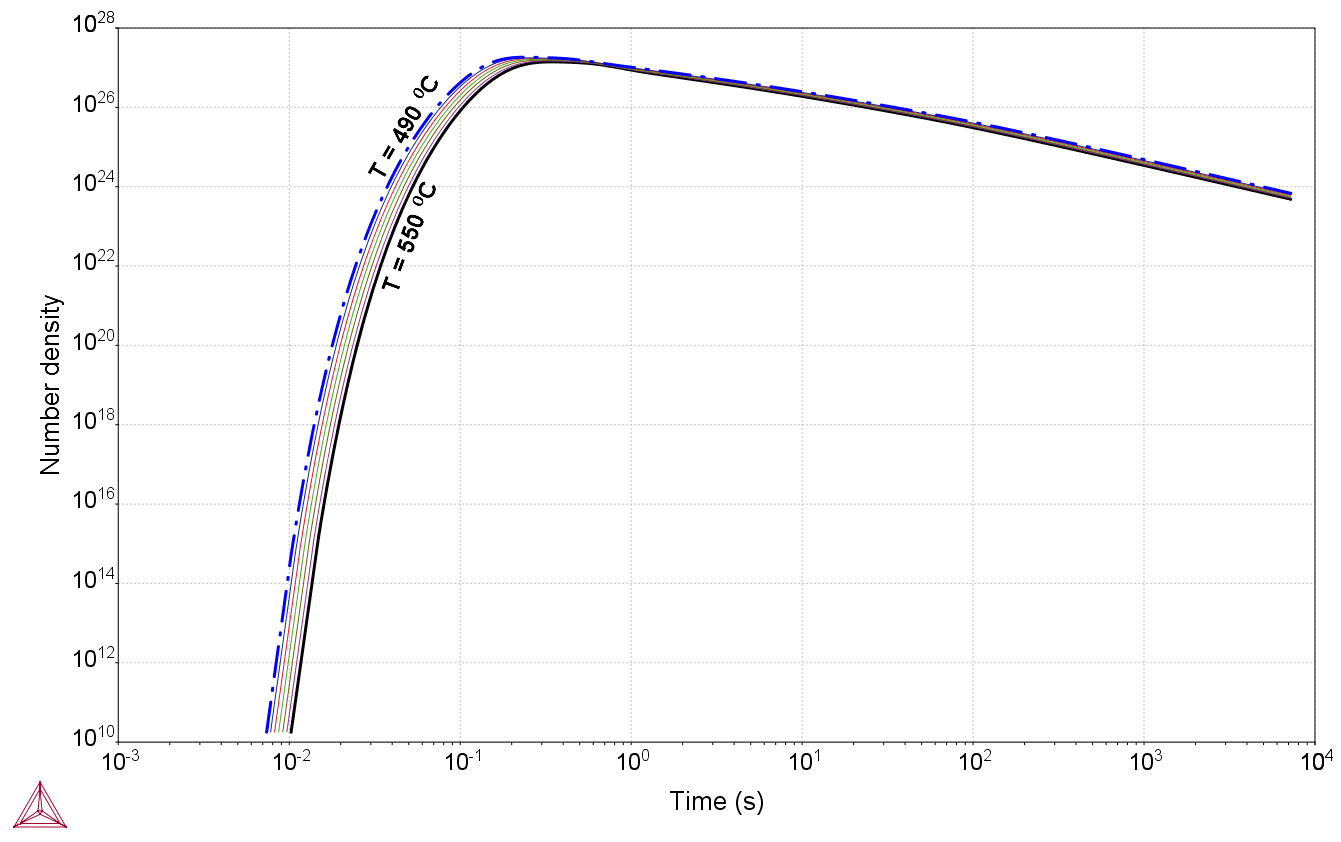}
\caption{Number density (of M3SI (Fe$_3$Si)) vs Time during isothermal annealing at various annealing temperatures.}
\label{Number_density_Time_Initial}
\end{figure}

Figure \ref{Number_density_Time_Initial} shows the plot of variation of number density of M3SI (Fe$_3$Si) nanocrystals  during isothermal annealing at various annealing temperatures (490-550 \degree C) for 2 hour holding time. Here too, we can observe a slight decrease in number density with increase in temperature. Thus lower number of precipitates with increase in annealing temperatures can be another reason for lower volume fraction at elevated temperatures (Figure \ref{Volume_fraction_Time_Initial} and Table \ref{tab_Mean_diameter_Time} reported in Sec.~\ref{res_Simulating_Fe3Si}). Reported number density is in accordance with reported work on FINEMET alloys \cite{dia_clavaguera2002crystallisation, dia_hono1999cu}.

\subsubsection{ Driving Force vs Time} \label{a_driving_force}
\begin{figure}[ht]
\centering
\includegraphics[width = 8.4 cm]{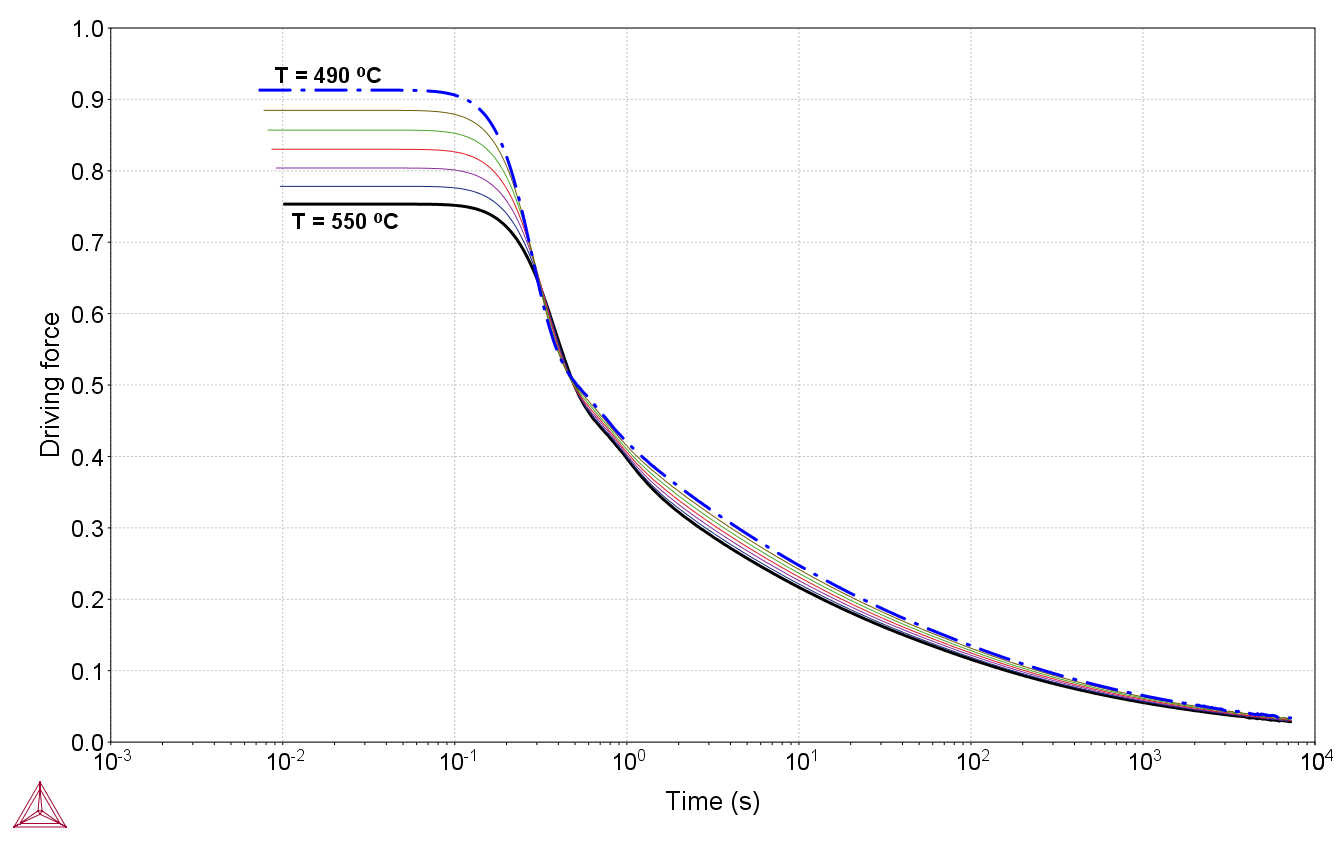}
\caption{Driving Force (of formation of M3SI (Fe$_3$Si) nanocrystal) vs. Time during isothermal annealing at various annealing temperatures.}
\label{Driving_force_Time_Initial}
\end{figure}

Figure \ref{Driving_force_Time_Initial} shows variation of driving force of formation of M3SI (Fe$_3$Si) nanocrystal during isothermal annealing at various annealing temperatures (490-550 \degree C) for 2 hour holding time. Here, too we can observe that driving force decreases with increase in temperature. This can explain the decrease in nucleation rate, number density and thus decrease in volume fraction with increase in isothermal annealing temperature (Figure \ref{Volume_fraction_Time_Initial} and Table \ref{tab_Mean_diameter_Time} reported in Sec.~\ref{res_Simulating_Fe3Si}).

\subsubsection{Matrix Composition}\label{a_matrix_comp}
In this section, matrix composition of Niobium, Boron and Copper obtained after isothermal annealing at a set of annealing temperatures (490 \degree C - 550 \degree C) for 3600, 5400 and 7200 seconds holding time has been tabulated in Table \ref{tab_Matrix_Nb}. Figure \ref{Matrix_composition_Time_Initial} in Sec.~\ref{res_Simulating_Fe3Si} shows the variation of matrix composition (mole \%) during isothermal annealing at 490 \degree C and 550 \degree C for Silicon, Niobium, Boron and Copper.

%%\begin{table}[H]
%%  \centering
%%  \caption{Matrix Composition of Niobium during isothermal annealing at various annealing temperatures for 3600, 5400 and 7200 seconds holding time}

%%\begin{tabular}{lllll}
%%%\toprule
%%\hline
%%\hline
%%    Temp. ($^{\circ}$C) & \multicolumn{3}{c}{Niobium (mole \%)} \\
%%    \hline
%%			    & 3600 s & 5400 s & 7200 s  \\
%%\hline
%%    %    \midrule
%%   490 & 7.915 & 7.922 & 7.928 \\
%%    %    \midrule
%%    500 & 7.868 & 7.875 & 7.883 \\
%%    %    \midrule
%%    500 & 7.868 & 7.875 & 7.883 \\
%%    %    \midrule
%%    520 & 7.767 & 7.778 & 7.785 \\
%%    %    \midrule
%%    530 & 7.712 & 7.725 & 7.734 \\
%%    %    \midrule
%%    540 & 7.660 & 7.675 & 7.681 \\
%%    %    \midrule
%%    550 & 7.605 & 7.615 & 7.625 \\
%%    %    \bottomrule
%%\hline
%%\end{tabular} \label{tab_Matrix_Nb}%
%%\end{table}%

\begin{table*}[ht]
  \centering
  \caption{Matrix Composition of Niobium during isothermal annealing at various annealing temperatures for 3600, 5400 and 7200 seconds holding time}

\begin{tabular}{lllllllllllllll}
%\toprule
\hline
\hline
    Temp. ($^{\circ}$C) & \multicolumn{5}{c}{Niobium (mole \%)} & \multicolumn{5}{c}{Boron (mole \%)} & \multicolumn{3}{c}{Copper (mole \%)} \\
    \hline
			    & 3600 s & 5400 s & 7200 s & & & 3600 s & 5400 s & 7200 s & & & 3600 s & 5400 s & 7200 s \\
\hline
    %    \midrule
    490 & 7.915 & 7.922 & 7.928 & & & 18.210 & 18.228 & 18.241 & & & 2.647 & 2.649 & 2.651 \\
    %    \midrule
    500 & 7.868 & 7.875 & 7.883 & & & 18.103 & 18.118 & 18.136 & & & 2.631 & 2.633 & 2.636 \\
    %    \midrule
    510 & 7.868 & 7.875 & 7.883 & & & 17.994 & 18.014 & 18.024 & & & 2.615 & 2.618 & 2.620 \\
    %    \midrule
    520 & 7.767 & 7.778 & 7.785 & & & 17.869 & 17.896 & 17.911 & & & 2.597 & 2.601 & 2.60 \\
    %    \midrule
    530 & 7.712 & 7.725 & 7.734 & & & 17.744 & 17.774 & 17.794 & & & 2.579 & 2.583 & 2.586 \\
    %    \midrule
    540 & 7.660 & 7.675 & 7.681 & & & 17.6231 & 17.6579 & 17.672 & & & 2.561 & 2.566 & 2.568 \\
    %    \midrule
    550 & 7.605 & 7.615 & 7.625 & & & 17.498 & 17.520 & 17.543 & & & 2.543 & 2.546 & 2.550 \\
    %    \bottomrule
\hline
\end{tabular} \label{tab_Matrix_Nb}%
\end{table*}%

\bibliographystyle{elsarticle-num}
\section*{References}

%\bibliography{FINEMET}

\end{document}